\documentclass[useAMS,usenatbib,usegraphicx,onecolumn]{mn2e}
\usepackage{times}

\newcommand{\bp}{\bmath{p}}
\newcommand{\bP}{\bmath{P}}
\newcommand{\bx}{\bmath{x}}
\newcommand{\bX}{\bmath{X}}
\newcommand{\bV}{\bmath{V}}
\newcommand{\mtc}{(\mt-m_C)}
\newcommand{\mb}{m_{\rm b}}
\newcommand{\mt}{m_{\rm t}}
\newcommand{\map}{m_{(A+p)}}
\newcommand{\mbp}{m_{(B+p)}}
\newcommand{\mcp}{m_{(C+p)}}
\newcommand{\bs}{\bmath{s}}
\newcommand{\ZP}{\sum{}\bmath{P}_j}
\newcommand{\Zp}{\sum{}\bmath{p}_j}
\newcommand{\Z}{\sum_{j=1}^N}
\newcommand{\Zm}{\sum{}m_j}

\newcommand{\Zmsp}{\sum_jm_{SP_j}}
\newcommand{\Zmp}{\sum_jm_{P_j}}
\newcommand{\Zmxa}{\sum_jm_{SA_j}\bmath{x}_{SA_j}}
\newcommand{\Zmxb}{\sum_jm_{SB_j}\bmath{x}_{SB_j}}
\newcommand{\Zmxc}{\sum_jm_{SC_j}\bmath{x}_{SC_j}}
\newcommand{\Zmxsp}{\sum_jm_{SP_j}\bmath{x}_{SP_j}}
\newcommand{\Zmxp}{\sum_jm_{P_j}\bmath{x}_{P_j}}
\newcommand{\ZmX}{\sum{}m_j\bmath{X}_j}

\newcommand{\Hk}{H_{\rm Kep}}
\newcommand{\Hi}{H_{\rm Int}}
\newcommand{\Hj}{H_{\rm Jump}}

\newcommand{\Gij}{\sum_{{^{j=1} _{j\neq i}}}^N \frac{Gm_j}{R_{ij}^3} (\bX_i - \bX_j)}

\newcommand{\lf}{\nonumber\\}
\newcommand{\cube}[1]{\frac{#1}{|#1|^3}}
\title[Triple system stability]{Planetary Stability Zones in Hierarchical Triple Star Systems}
\author[P. E. Verrier and N. W. Evans] 
{P. E. ~Verrier$^1$\thanks{E-mail: pverrier@ast.cam.ac.uk (PEV); nwe@ast.cam.ac.uk (NWE)} 
and N. W. ~Evans$^1$\footnotemark[1]\\
$^1$Institute of Astronomy, University of Cambridge, Madingley Road, Cambridge, CB3 0HA, United Kingdom}
\begin{document}

\date{This draft 2007 February 1}

\pagerange{\pageref{firstpage}--\pageref{lastpage}} \pubyear{0000}

\maketitle

\label{firstpage}


\begin{abstract}
A symplectic integrator algorithm suitable for hierarchical triple
systems is formulated and tested. The positions of the stars are
followed in hierarchical Jacobi coordinates, whilst the planets are
referenced purely to their primary. The algorithm is fast, accurate
and easily generalised to incorporate collisions. There are five
distinct cases -- circumtriple orbits, circumbinary orbits and
circumstellar orbits around each of the stars in the hierarchical
triple -- which require a different formulation of the symplectic
integration algorithm.

As an application, a survey of the stability zones for planets in
hierarchical triples is presented, with the case of a single planet
orbiting the inner binary considered in detail.  Fits to the inner and
outer edges of the stability zone are computed. Considering the
hierarchical triple as two decoupled binary systems, the earlier work
of Holman \& Wiegert on binaries is shown to be applicable to triples,
except in the cases of high eccentricities and close or massive stars.
Application to triple stars with good data in the multiple star
catalogue suggests that more than 50 \% are unable to support
circumbinary planets, as the stable zone is non-existent or very
narrow.
\end{abstract}




\section{Introduction}
\label{sec:intro}

It is well known that many of the stars in the solar neighbourhood
exist in multiple systems. As the number of planetary surveys
increases, planets are regularly being found not only in single star
systems, but binaries and triples as well. For example, recently a hot
Jupiter has been claimed in the triple system HD 188753 (Konacki 2005;
but see also Eggenberger et al. 2007), and \citet{DB07} lists 33
binary systems and 2 other hierarchical triples known to harbour
exoplanets. As the majority of work on planetary dynamics has been for
single star systems, the dynamics of bodies in these multiple stellar
systems is of great interest. At first sight, it might not seem likely
to expect long term stable planetary systems to exist in binary star
systems, let alone triples, but numerical and observational work is
starting to show otherwise.

In recent years, much study has been devoted to the stability of
planets and planetesimal discs in binary star systems. There are
several investigations of individual systems (e.g. work on $\gamma$
Cephei by \citealt{Dv03}, \citealt{Ha06} and \citealt{VE06}) as well
as substantial amounts of work on more general limits for stability of
smaller bodies in these systems.  Notably, \citet{HW99} approach this
problem by using numerical simulation data to empirical fit critical
semimajor axes for test particle stability as functions of binary mass
ratio and eccentricity.  These general studies are of great use in the
investigation of observed systems and their stability properties,
giving an effective and fast method of placing limits on stability in
the large parameter space created by observational uncertainties.

The aim of this work is to investigate test particle stability in
hierarchical triple star systems, and to see if any similar boundaries
can be defined for them.  To do this, a new method for numerically
integrating planets in such systems is presented, following the ideas
for binary systems presented by \citet{CQDL02}.

Although there have been empirical studies of the stability of
hierarchical star systems themselves, there do not appear to have been
studies of small bodies in such systems. There is a great deal of
literature concerning periodic orbits in the general three and four
body problems (see e.g. \citealt{Sz67}), but these are almost entirely
devoted to considerations of planetary satellites in single star
systems, for example satellite transfer in the Sun-Earth-Moon systems.
Also, while periodic orbits prove that stable solutions exist in these
problems, they are of little practical use in determining general
stability limits.

The problem of planetary orbits in triple systems is more complex than
for those in binary systems, as there are many different orbital
configurations possible relative to the three stars. These are
considered in Section~\ref{sec:orbits}, and a method for classifying
them is described in order to simplify the discussion in this paper.
The derivation of a method to numerically integrate such planets is
then given in Section~\ref{sec:maths}.  In Section~\ref{sec:stats} is
a brief overview of the statistical properties of known triple
systems, as a basis for deciding the parameters of the systems used in
the numerical simulations. The results of numerical investigations
into stability properties are then presented in Section~\ref{sec:sp}
for one of the possible orbital types.


\section{The Types of Stellar and Planetary Orbits}
\label{sec:orbits}

The triple systems studied here are all hierarchical in nature.  That
is, they can all be considered to be a close binary orbited by a
distant companion -- in effect, an inner binary pair and an outer
binary pair. Here, the inner binary stars are labelled as A and B,
with A being the more massive, and the distant companion star as C.
There are however three possible ways to pair the three stars into the
hierarchy described above. \citet{EK95} define the hierarchy by
requiring that the instantaneous Keplerian orbits of the inner and
outer pair are bound, that the outer binary has a longer period than
the inner binary and that the ratio of the two periods is the largest
of the three possible pairings of the stars. This definition is
adopted here.

Although other, non-hierarchical, types of motion are possible for
triple stars (see for example \citealt{Sz77}) and observed (see for
example the trapezium systems listed in \citealt{To97}), they are not
considered here. It is an open question whether non-hierarchical
systems can sustain long-term stable planets.

Next some system of classifying planetary orbits is needed. The
orbital motion of small particles under the influence of two or three
masses has been studied to some extent through periodic orbits of the
three and four body problems, as mentioned in the Introduction. As a
result of this many families and classification schemes exist.
\citet{Sz67} gives a good review, mentioning for example the (a) to
(r) types designated for the Copenhagen problem, dependent on the
nature and location of the particles motion in both inertial and
corotating frames. Many other features have been used to describe
orbital families, for example symmetry \citep{Br04} or a parameter
used to generate the orbital family. However, as mentioned by
\citet{Sz67}, there is no overall method, and the descriptions would
not be applicable to non-periodic orbits. 

\begin{figure*}
\centering
\includegraphics[width=\textwidth]{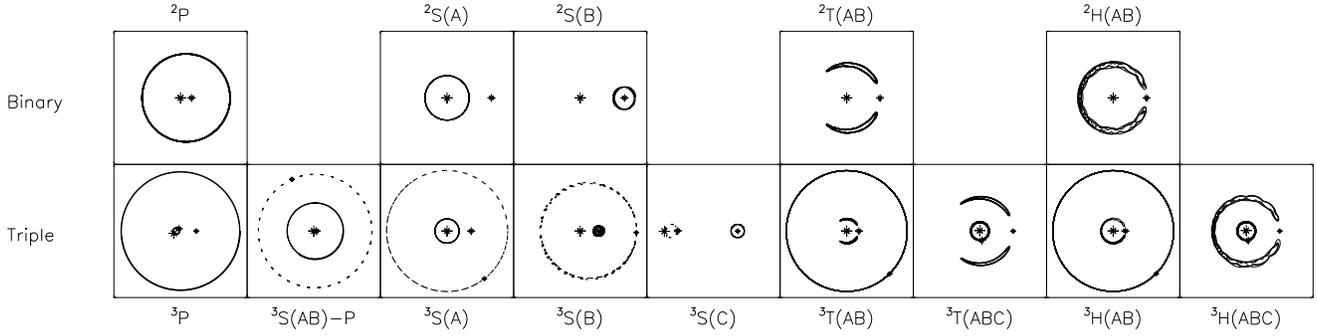}
\caption{
  Basic planetary orbit types in multiple star systems as described in
  Section~\ref{sec:orbits}. The top row shows a binary system, in a
  frame corotating with the stars. The bottom row shows a triple
  system, in a frame corotating with either the inner or outer binary
  as appropriate. Stellar orbits are shown with dashed lines and stars
  marked with an asterisk. Planetary orbits are shown with a solid
  line. Note that these plots show real data from numerical
  simulations generated using a standard Bulirsch-Stoer integrator
  \citep{NR} for the tadpole and horseshoe orbits (based on initial
  conditions from \citealt{MD00}) and the integration scheme presented
  in this paper for the others.}
\label{fig:types}
\end{figure*}

Orbital types in binary systems have generally been classified into
three main groups, those around a single star (circumstellar), those
around both stars (circumbinary) and those in the middle ground as it
were, coorbiting with one of the stars i.e. librating about the
triangular Lagrangian points, similar to Trojan asteroids in the Solar
System. A convenient labelling of these was given by \citet{Dv84}, who
designated planets as either S-orbits (satellite), P-orbits (planet),
or L-orbits (librator) for circumstellar, circumbinary and coorbital
respectively.  These ideas can be extended to hierarchical triple
systems fairly simply. It is clear that there will be three types of
circumstellar motion, one for each star, and as such orbits like this
can be labelled S(A), S(B) and S(C) for their primary. The
circumtriple orbit about the centre of mass of all three can be
identified as a P orbit and labelled as such in analogy to Dvorak's
scheme. Finally, planets which orbit the centre of mass of the inner
binary are circumbinary but also share characteristics with the
satellite orbit, and can be labelled with the combination S(AB)-P.
These orbital types are shown in Figure~\ref{fig:types} along with the
binary cases for comparison, and listed in Table~\ref{tab:types}. A
superscript 2 has been given to those in binary systems and a 3 for
those in triple systems for clarity.

Although these are the only types of motion studied here, for
completeness the coorbital types are also be included. This type of
motion can occur for both the inner and outer binary, and labels (AB)
and (ABC) can be used to designate this. Instead of the single
L-orbit type they can be broken up into T and H orbits to indicated
tadpole (about one of the triangular Lagrangian points) or horseshoe
(about both triangular points and the $L_3$ collinear point). These
are again illustrated in Figure~\ref{fig:types} and included in
Table~\ref{tab:types}.

These orbits are defined to be bound relative to their focus and
hierarchical in the same sense as the definition given for the stellar
system. A particle that orbits outside the extent of the inner binary
but has a bound orbit with respect to star A and the binary centre of
mass would be an S(AB)-P orbit and not an S(A) orbit.

This method of labelling clearly and concisely designates the exact
nature of a planetary orbit, and extends a system already in use for
binary stars. It can also be easily applied to systems with other
levels of hierarchy, for example if star C was replaced with another
close binary pair C and D additional classes S(D) and S(CD)-P would be
possible, and the outer P type could be relabelled P(ABCD).

\begin{figure}
\begin{center}
\includegraphics[width=0.7\textwidth]{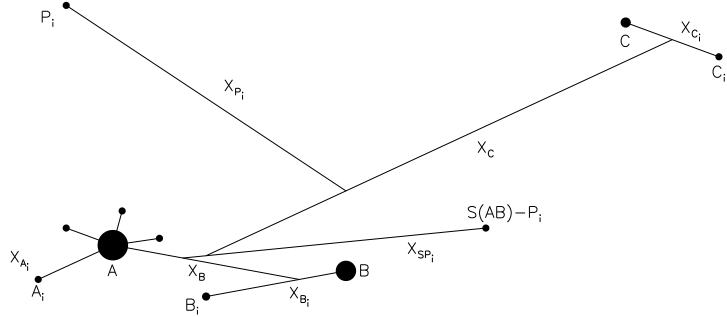}
\end{center}
\caption{
  The coordinate system used for the symplectic integrator. Planets
  are all taken as relative to their primary, whilst each
  planetary-stellar subsystem barycentre is referred to the centre of
  mass of the preceding objects.}
\label{fig:coord}
\end{figure}

\begin{table}
\centering
\caption{
  Labels for the basic planetary orbital types in multiple star systems
  as defined in this work and compared to those of \citet{Dv84}.
}
\begin{tabular}{l|lll}
Orbit                  & Triple               & Binary        & Dvorak      \\
description            & system               & system        &  (1984)     \\
\hline
Circumstellar          & S(A)                 & S(A)          &   S         \\
                       & S(B)                 & S(B)          &  --         \\
                       & S(C)                 & --            &  --         \\
Circumbinary           & S(AB)-P              & P(AB)         &   P         \\
Circumtriple           & P or P(ABC)          &  --           &   --        \\
Coorbital with binary  & T(AB)                & T(AB)         &   L         \\
                       & H(AB)                & H(AB)         &   --        \\
Coorbital with triple  & T(ABC)               &  --           &   --        \\
                       & H(ABC)               &  --           &   --        \\
\end{tabular}
\label{tab:types}
\end{table}


\section{A Symplectic Integrator for Hierarchical Triples}
\label{sec:maths}

Given the number of simulations that are needed to accurately describe
the stability of general triple systems, an accurate and fast method
of numerically integrating the equations of motion is required.
Methods such as Bulirsch-Stoer are reasonably accurate, but very slow.
Instead, a symplectic method is favoured, as it is fast and shows
excellent energy conservation. 

A symplectic integration method for planetary systems was introduced
by \citet{WH91}. Here, the Hamiltonian of the system is split into a
dominant Keplerian part and smaller interaction terms, all
analytically integrable on their own.  A leapfrog scheme is then
applied to evolve the system forwards by one timestep. If $H_K$ is the
Keplerian Hamiltonian and $H_I$ the smaller interaction terms, then to
move forward by one step of $dt$ the system is evolved by $H_I$ for
$dt/2$, $H_K$ for $dt$ and finally by $H_I$ for $dt/2$ again. This is
a 2nd order method.

There have been two symplectic integrators derived for planets in
multiple star systems, by \citet{CQDL02} and \citet{Be03}
respectively.  \citet{Be03} uses modified Jacobi coordinates for
hierarchical systems of any multiplicity. These `hierarchical Jacobi
coordinates' account for the separate substructures in the
hierarchy. Following the example given by \citet{Be03}, if a system
consists of four stars in two separate binaries orbiting the systems
barycentre, then each pair is assigned Jacobi coordinates within their
subsystems, and then Jacobi coordinates are applied to the two
binaries.  \citet{CQDL02} uses a different modification.  Here, the
stars are in these hierarchical Jacobi coordinates, but planets are
referenced purely to their primary. This permits close encounters to
be implemented within each planetary system, but requires the small
interaction Hamiltonian to be split into two separate parts. The
system is evolved using leapfrog still, and the method is 2nd order so
long as the order the two interaction terms is symmetric.  Both
symplectic integrators can be applied to the problem here, but that of
\citet{CQDL02} is favoured due to the ease of implementing close
encounters. Note that, in the limit of test particles only or a single
planet, they are identical coordinate systems.

This method requires some work to be extended to a triple system.
First the coordinate system needs to be defined, as shown in
Figure~\ref{fig:coord}.  S(A) planets are taken relative to star A,
S(B) to star B and S(C) to star C. The barycentre of star B and its
planetary system is then referred to the barycentre of star A and its
planets (similar to \citet{Be03}'s hierarchical coordinates). S(AB)-P
planets are taken as relative to the centre of mass of the binary and
all planets therein.  The barycentre of the S(C) planets and their
star is taken as relative to the binary and its circumstellar and
circumbinary planetary systems. Finally, P type planets are referred to
the centre of mass of all the other subsystems. Following the notation
of \citet{CQDL02}, the transformed coordinates $\bX$ are related to
the inertial coordinates $\bx$ by
\begin{eqnarray}
\mt\bX_A   &=& m_A\bx_A + m_B\bx_B + m_C\bx_C    
                   + \Zmxa + \Zmxb    
                   + \Zmxc + \Zmxsp    
                   +  \Zmxp  \lf 
\bX_B      &=& \frac{m_B\bx_B + \Zmxb}{\mbp}    
                - \frac{m_A\bx_A + \Zmxa}{\map} \lf
\bX_C      &=& \frac{m_C\bx_C + \Zmxc}{\mcp}     
                   - \frac{m_A\bx_A + \Zmxa +              
                           m_B\bx_B + \Zmxb + \Zmxsp}               
                                  {\mb+\Zmsp} \lf
\bX_{SA_i} &=& \bx_{SA_i} - \bx_A \lf
\bX_{SB_i} &=& \bx_{SB_i} - \bx_B \lf
\bX_{SC_i} &=& \bx_{SC_i} - \bx_C \lf
\bX_{SP_i} &=& \bx_{SP_i}    
                          - \frac{m_A\bx_A + \Zmxa          
                                 +m_B\bx_B + \Zmxb}{\mb} \lf
\bX_{P_i}  &=& \bx_{P_i} - \frac{\mt\bX_A - \Zmxp}{\mt - \Zmp}
\end{eqnarray}
where subscripts $A$, $B$, $C$ label the stars and $SA_i$, $SB_i$,
$SC_i$, $SP_i$ and $P_i$ label the planets.  $\mt$ is the total mass
of the system and $\mb$ is the mass of the inner binary stars and
their planets. To derive the full system of equations is somewhat
involved, but individually (i.e. for one orbital type only) they are
readily obtainable. For example, using the method of \citet{CQDL02},
the conjugate momenta for a hierarchical triple with $N$ S(AB)-P
planets only are
\begin{eqnarray}
\bP_A &=& \bp_A + \bp_B + \bp_C  + \Zp \lf
\bP_B &=& \bp_B - \frac{m_B}{\mb}(\bp_A+\bp_B) \lf
\bP_i \,\, &=&  \bp_i - \frac{m_i}{\mtc}(\bp_A + \bp_B + \Zp) \lf
\bP_C &=& \bp_C - \frac{m_C}{\mt}(\bp_A + \bp_B + \bp_C + \Zp) 
\end{eqnarray}
where here $\mb$ is the mass of the inner binary only. The transformed
Hamiltonian is split as follows
\begin{eqnarray}
H \quad \,\,&=& \Hk + \Hi + \Hj \lf
\Hk &=& \frac{\bP_{B}^{2}}{2\mu_{b}} + \frac{\bP_{C}^{2}}{2\mu_{t}}
               - \frac{G\mu_{b}\mb}{R_B} - \frac{G\mu_{t}\mt}{R_C}    
               + \Z \left( \frac{\bP_{i}^2}{2m_i}-\frac{G\mb m_i}{R_i} \right)\lf
\Hi &=& -\Z \sum_{j>i} \frac{Gm_im_j}{R_{ij}}    
            + G\mb m_{C}\Bigg( \frac{1}{R_C} 
            -            \frac{m_A}{|\mb\bX_C+m_B\bX_B+\mb\bs|}    
            -            \frac{m_B}{|\mb\bX_C-m_A\bX_B+\mb\bs|}\Bigg)\lf
& &{}       + G\mb\Z m_i \Bigg( \frac{1}{R_i}    
            -  \frac{m_A}{|\mb\bX_i+m_B\bX_B|}    
            -  \frac{m_B}{|\mb\bX_i-m_A\bX_B|} \Bigg)\lf
& &{}       + Gm_C \Z m_i \left(\frac{1}{R_C} - \frac{1}{|\bX_C -
              \bX_i +\bs|}\right)\lf
\Hj &=& \frac{1}{2\mb}\left|\ZP\right|^2
\end{eqnarray}
where $\mu_{b} = m_Am_B/\mb$ and $\mu_{t} = m_C\mtc / \mt$ are the
reduced masses of the inner and outer binaries respectively and $\bs =
\ZmX/\mtc$. The Hamiltonian $\Hk$ represents the Keplerian
motion of the stellar orbits and the Keplerian motion of the planets
about a fixed point. The recommended method by \citet{WH91} is to use
the f and g functions of \citet{Da88} to evolve the system under this
term.

The Hamiltonian $\Hi$ contains interaction terms dependent only on
position. Hamilton's equations can be used to derive the accelerations
on each object due to this term, and these can be analytically integrated
to evolve the system over the interval $dt$. These accelerations are
\begin{eqnarray}
\frac{d\bmath{V}_{B}}{dt} &=& -G\mb m_A\Z m_i \Bigg(\cube{\mb\bX_i+m_B\bX_B}   
                              -               \cube{\mb\bX_i-m_A\bX_B}\Bigg)\lf
& &{}                         - G\mb m_Am_C \Bigg( \cube{\mb\bX_C +m_B\bX_B +\mb\bs}    
                              -               \cube{\mb\bX_C -m_A\bX_B +\mb\bs}\Bigg)\lf
\frac{d\bmath{V}_{C}}{dt} &=& \frac{G\mtc\bX_C}{R_C^3}      
                             - G\Zm\cube{\bX_C - \bX_j +\bs}  \lf
& &{}                          - G\mb^2\Bigg( m_A\cube{\mb\bX_C+m_B\bX_B +\mb\bs}    
                                +        m_B\cube{\mb\bX_C -m_A\bX_B +\mb\bs}\Bigg) \lf
\frac{d\bmath{V}_{i}}{dt} &=&  Gm_C \cube{\bX_C - \bX_i +\bs} 
                                -\frac{Gm_C}{\mtc}\Z m_j\cube{\bX_C - \bX_j +\bs} - \Gij \lf
& &{}                           -\frac{G\mb^2m_C}{\mt - m_C} \Bigg(m_A \cube{\mb\bX_C +m_B\bX_B +\mb\bs}   
                                +                            m_B \cube{\mb\bX_C -m_A\bX_B +\mb\bs}\Bigg)\lf
& &{}                           + G\mb \Bigg( \frac{\bX_i}{R_i^3}    
                                -      m_A\mb\cube{\mb\bX_i+m_B\bX_B}    
                                -      m_B\mb\cube{\mb\bX_i-m_A\bX_B} \Bigg)
\end{eqnarray}
Note that in these equations $\bV$ is a pseudo-velocity and equal to
the transformed momenta divided by the mass of the objects they
describe. That is, $\bV_B$ is $\bP_B / \mb$ and so on. The final
Hamiltonian is $\Hj$ and gives
\begin{eqnarray}
\frac{d\bX_{i}}{dt} = \Z \frac{\bP_j}{\mb}
\end{eqnarray}
A similar method can be used to obtain the corresponding equations for
the other four orbital types, and these are given in the Appendix.
Further details of the derivation for a more general system with more
than one orbital type are given in \citet{Ve08}. Close encounters can
be included in the same way as described by \citet{CQDL02}, as well as
variable timesteps, but are not implemented here.

The method outlined above was implemented separately for each
planetary type in a stand alone program {\sc Moirai}\footnote{The
Moirai are in Greek mythology the three Fates. Given the tradition of
naming planetary objects after gods and heros from classical
mythology, and that the stars in these systems will be the largest
influence on the orbital evolution of such objects, it seemed
appropriate to name them (and the code) after the goddesses that were
supposed to control the fate of men and gods alike.}, the testing of
which is also described in the Appendix and in \citet{Ve08}.


\section{Triple Star System Statistics}
\label{sec:stats}

\begin{figure}
\centering
\includegraphics[width=\textwidth]{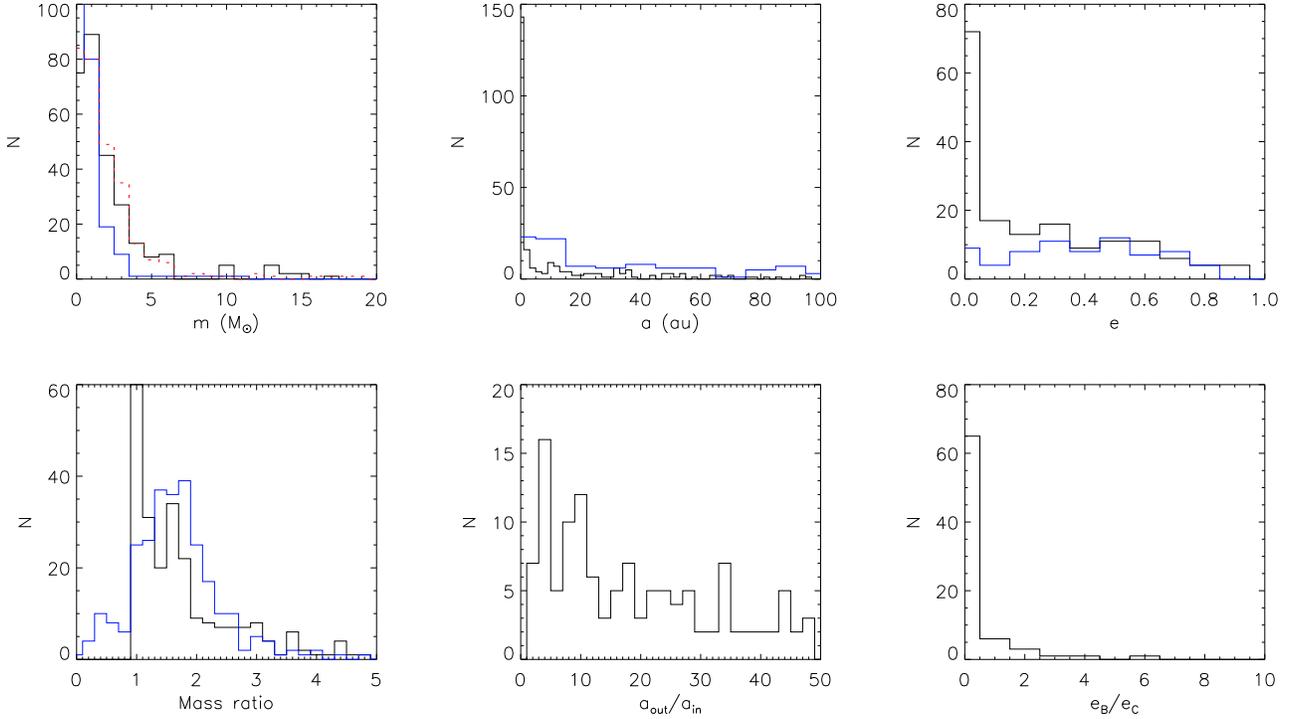}
\caption{
  The statistics of the orbits of stars in triple star systems,
  plotted using data from \citet{To97}. The top left panel shows a
  histogram of masses of each component (A black, B blue and C
  red-dashed), while the bottom left shows the mass ratios of the
  inner and outer binaries ($m_A/m_B$ black, $m_{bin}/m_C$ blue). The
  top middle panel shows the semimajor axes of the inner (black) and
  outer (blue) orbits, while the bottom middle panel shows the ratio
  of these. The top left panel shows the eccentricities of the inner
  (black) and outer (blue) orbits, and the bottom left panel shows the
  ratio of the two.}
\label{fig:stats}
\end{figure}

It is helpful to determine the statistical properties of known
hierarchical triple systems, so simulations can be run that are
comparable to real systems. The Multiple Star Catalogue \citep{To97}
contains 541 reasonably complete entries for such systems, and
histograms of this data are plotted in Figure~\ref{fig:stats}. This
catalogue is compiled from a variety of sources so contains a number
of biases from different observing methods, as discussed by the
author, but should be sufficient here. 

>From the histograms, it can be seen that the stellar masses tend to be
near solar, with the inner binary mass ratio being generally less than
2 and the outer between 1 and 3, although sometimes being larger than
the total mass of the binary. The semimajor axis of the inner binary
tends to be less than 2au and the outer, although often many hundreds
of astronomical units, generally concentrated around a few tens. There
seems to be a big step in the distribution at about 15au. The ratio of
the semimajor axes ranges but seems to be slightly peaked towards a
factor of 10 or less. The inner eccentricity is peaked around zero,
and a plot of the inner eccentricity as a function of semimajor axis
shows that this does not only occur for close inner binaries. The
outer eccentricity is fairly uniform from 0 to about 0.7. A plot of
the outer eccentricity as a function of semimajor axis also shows no
apparent trend.

\citet{To97} considers the catalogue to be complete to a distance of
10pc. However, this includes only 14 objects with all stellar orbital
elements known, so to determine if the data is representative it was
ranked by distance and a comparison made of the first 250 entries to
the entire sample. No major difference was apparent between the two,
so the sample can be considered to give a reasonable overview of
triple systems.

\section{Numerical Integrations: S(AB)-P orbits}
\label{sec:sp}


\begin{figure}
\centering
\includegraphics[width=\textwidth]{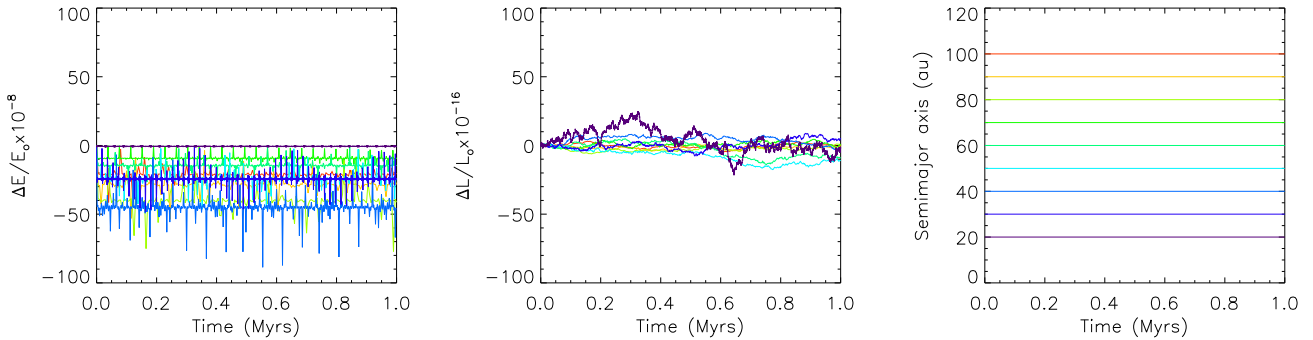}
\caption{
  Details of the sets of simulations with inner binary parameters
  $a_{B}=1$au, $e_{B}=0$, $m_A = m_B = 1 M_{\odot}$ and outer binary
  parameters $e_C = 0.6$, $m_C=4 M_{\odot}$ and $a_C =$ 20 to 100 au.
  The left and centre panels show the conservation of total energy and
  angular momentum for these eight simulations (note that the change
  in angular momentum is a factor of $10^{-8}$ smaller than that in
  energy, as in symplectic integration schemes angular momentum is
  conserved to machine precision). The right hand panel shows the
  variation of the semimajor axis of star C throughout the simulation.
  The different colours indicate the different initial values of the
  semimajor axis of star C, as apparent from the right panel. }
\label{fig:el}
\end{figure}


\begin{figure}
  \centering \includegraphics[width=\textwidth]{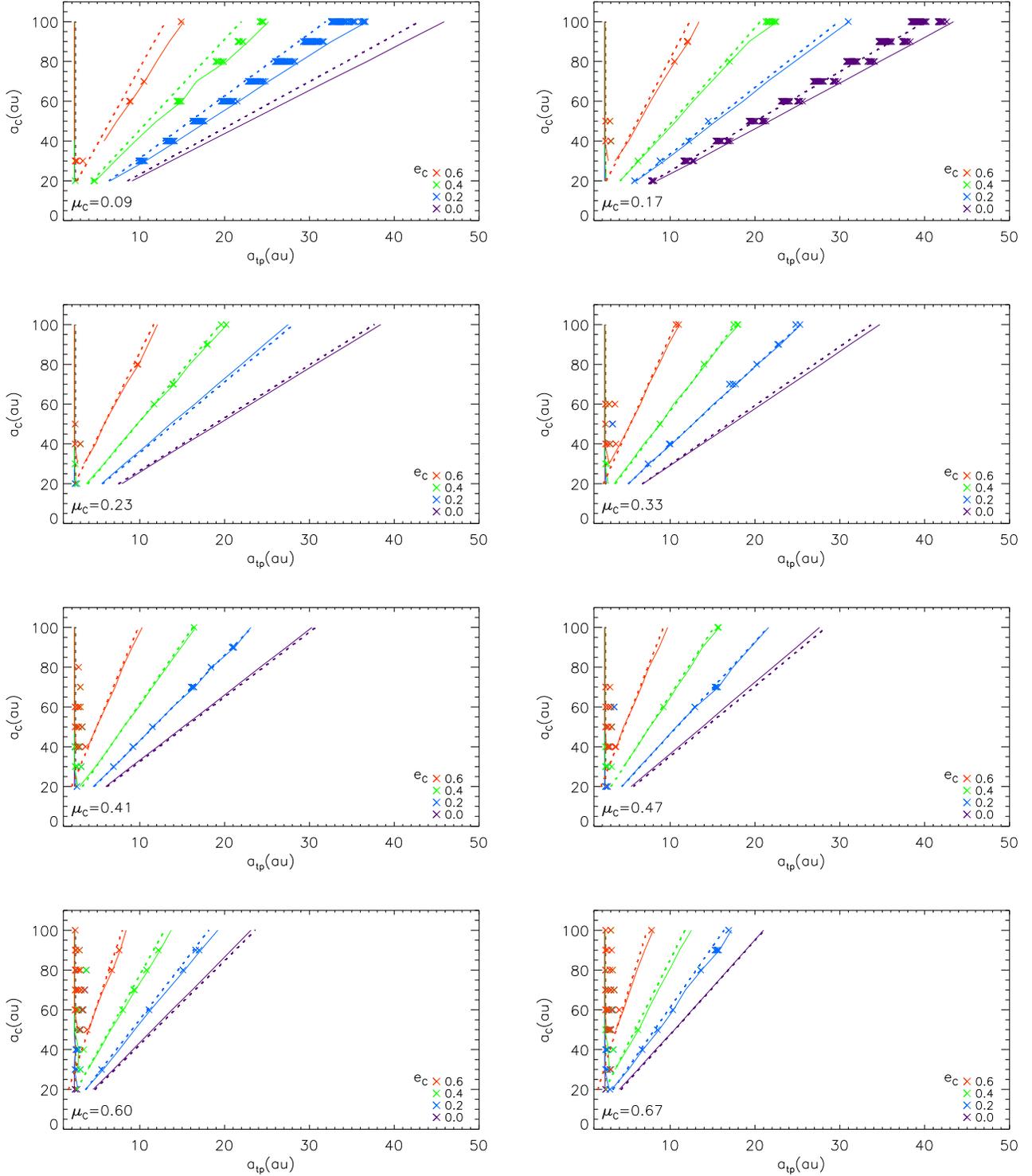}
\caption{
  The stability radii for S(AB)-P type test particles as a function of
  outer binary semimajor axis and eccentricity for eight different
  mass ratios. The colours indicate eccentricity of star C and the
  panels are labelled with the mass ratio. Solid lines show the first
  and last fully stable radii, and crosses show any unstable locations
  within this annulus. For comparison the location of the inner and
  outer critical semimajor axes predicted from \citet{HW99} are shown
  as dashed lines. }
\label{fig:sp_critout_i}
\end{figure}



\begin{figure}
\centering
\includegraphics[width=\textwidth]{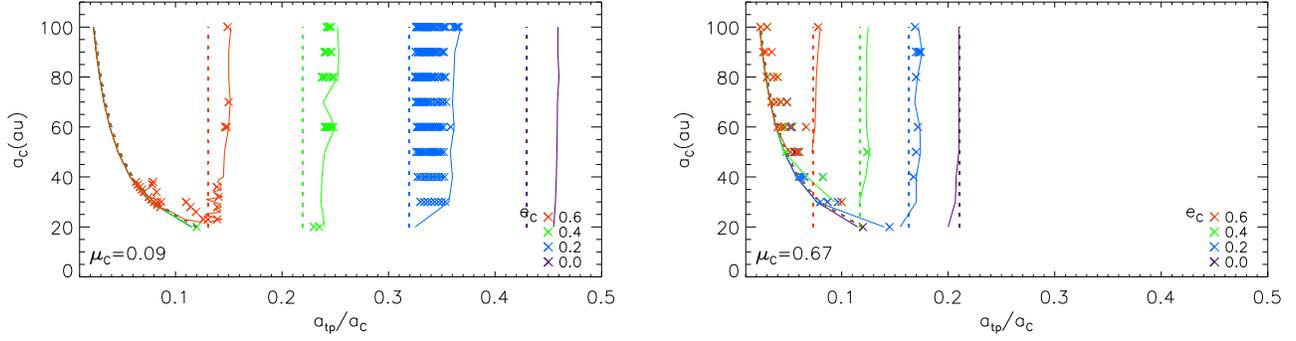}
\caption{
  The stability radii for S(AB)-P type test particles as a function of
  outer binary semimajor axis and eccentricity, this time scaled to
  the semimajor axis of star C, for the smallest and largest mass
  ratios studied. The symbols are as for Fig~\ref{fig:sp_critout_i}, and
  it can be seen that the location of the outer stability boundary
  scales with $a_C$.  }
\label{fig:sp_critout_o}
\end{figure}



\begin{figure}
\centering
\includegraphics[width=\textwidth]{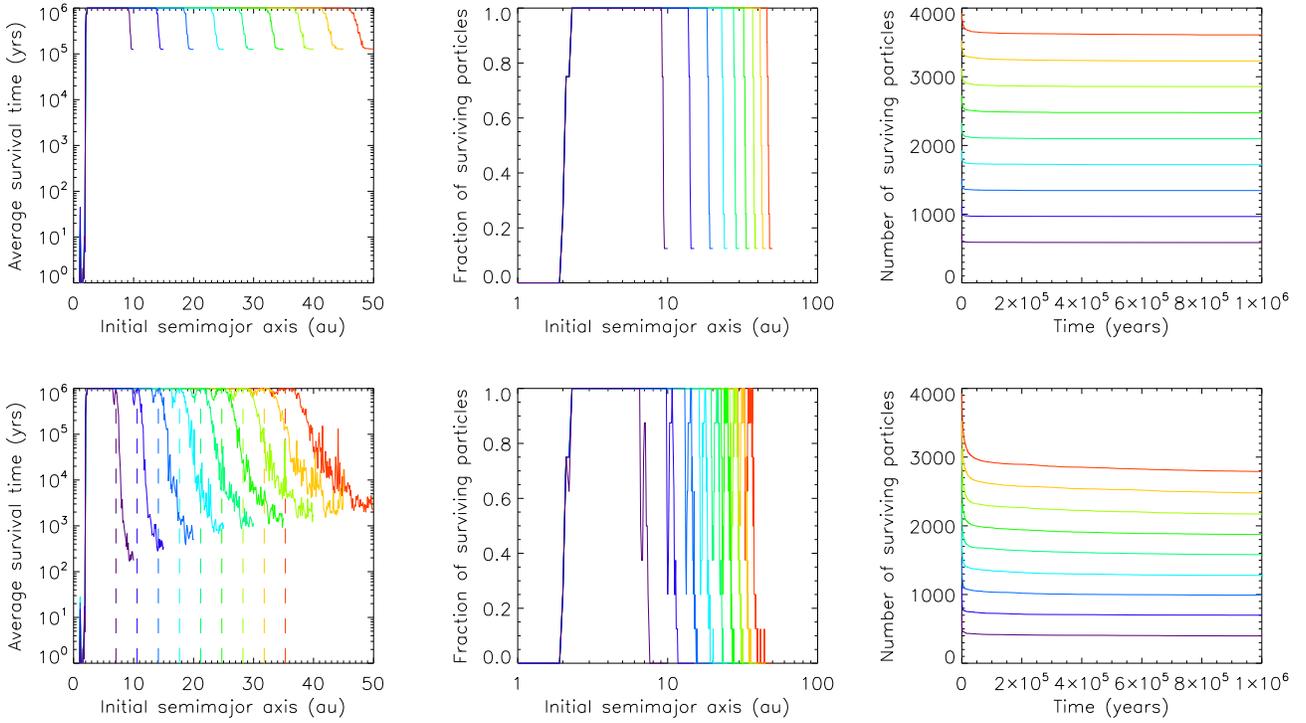}
\caption{
  Some examples of test particle evolution for simulations with
  $\mu_C=0.09$. Different colours indicate different values of $a_C$
  (see Figure~\ref{fig:el}). The left panels show average test
  particle survival time at each initial semimajor axis for $e_C=0.0$
  (top) and $e_C=0.2$ (bottom), indicating that for the zero
  eccentricity case the stable region is well defined, and that that
  the dynamics scale with $a_C$. The island around the 5:1MMR in the
  $e_C=0.2$ case is clearly visible, and the location of the resonance
  overlayed as a dashed line. The middle panel shows the fraction of
  test particles surviving at each initial semimajor axis and The
  right panel shows test particle decay rates. A fast clearing out of
  the unstable regions is seen.}
\label{fig:tpev01}
\end{figure}



\begin{figure}
\centering
\includegraphics[width=\textwidth]{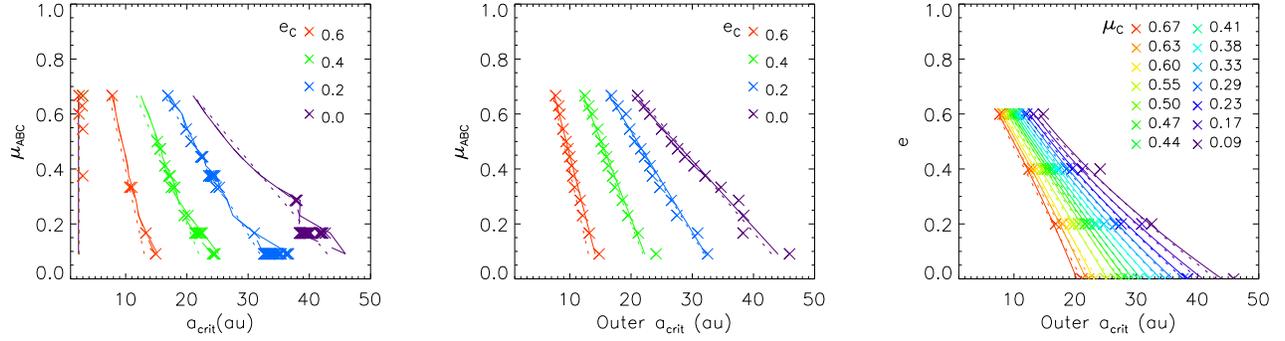}
\caption{ Test particle critical outer semimajor axes for the $a_C =
  100$ au cases: data and fits. In the left panel is the simulation
  data as a function of mass ratio, with symbols as before. The middle
  panel shows the 4 parameter fit to the data: now crosses correspond
  to the values to be fitted, the solid line is the fit and the dotted
  lines the results of \citet{HW99} for comparison. Note that the
  critical semimajor axis is taken as the innermost radii within the
  stable island not the line shown in the left panel. The right panel
  shows the fit as a function of eccentricity.  }
\label{fig:sp_crit_100}
\end{figure}



\begin{figure}
\centering
\includegraphics[width=\textwidth]{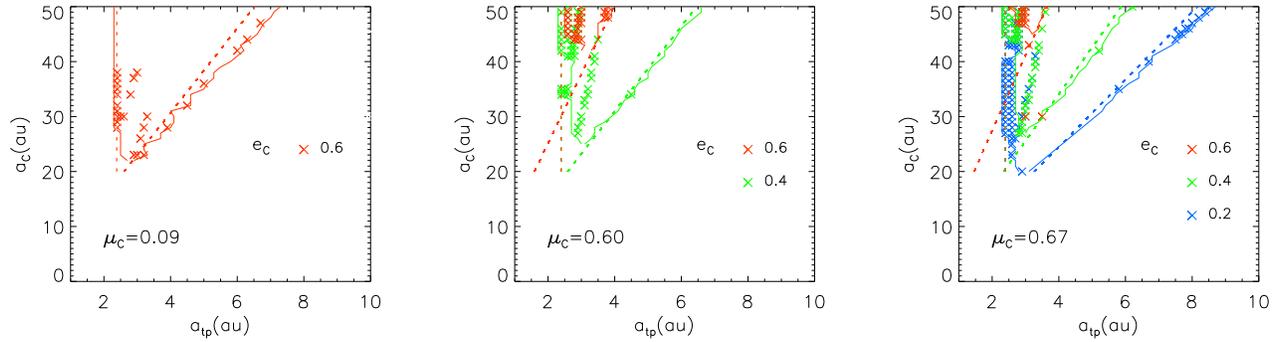}
\caption{
  Test particle stability as a function of initial semimajor axis, as
  for Figure~\ref{fig:sp_critout_i}, in more detail for the $\mu_C =
  0.09$, $\mu_C = 0.60$ and $\mu_C = 0.67$ simulations for various
  eccentricities of star C. Here additional simulations have been run
  for values of $a_C$ between 20 and 50 au in steps of 1 au. Symbols
  are as before: the solid lines are the inner and outer stable
  boundaries, crosses are unstable locations within these, and dotted
  lines the fit of \citet{HW99}.}
\label{fig:sp_crit_detail}
\end{figure}



\begin{figure}
\centering
\includegraphics[width=\textwidth]{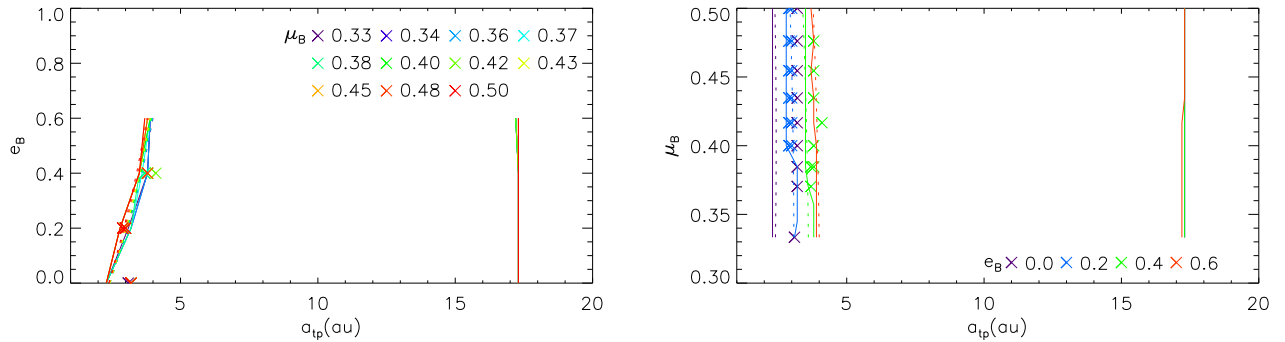}
\caption{
  The inner critical radius as a function of the inner binaries
  eccentricity and mass ratio. As for Figure \ref{fig:sp_critout_i}
  the solid lines show the first and last stable radius, crosses any
  unstable points between these two and the dashed lines the fit of
  \citet{HW99}. }
\label{fig:sp_crit_in}
\end{figure}


\begin{figure}
\centering
\includegraphics[width=\textwidth]{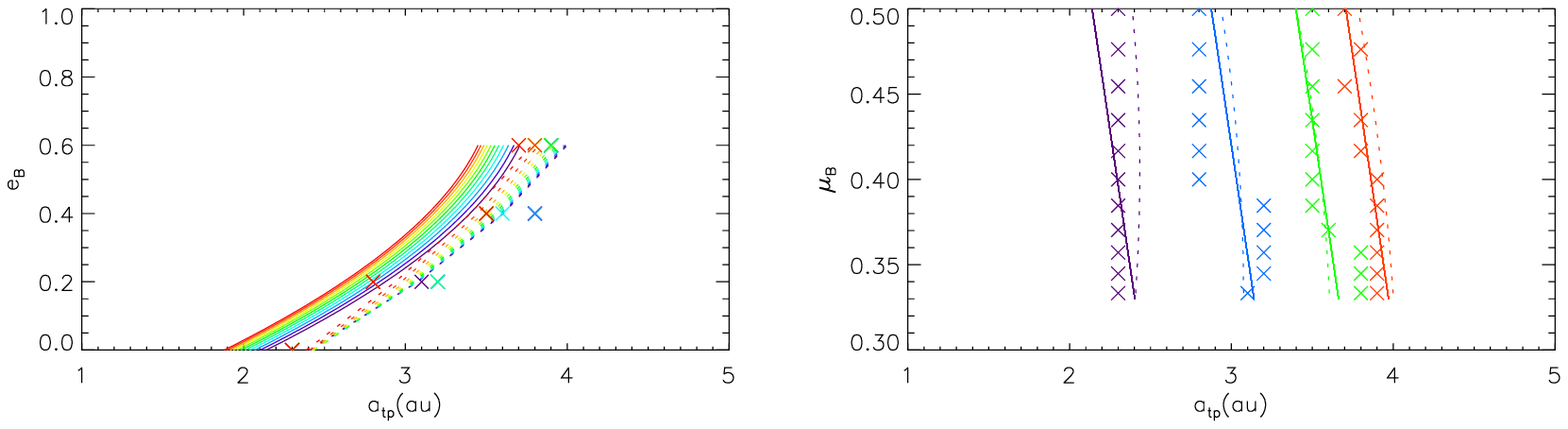}
\caption{
  The fit to the inner critical radius as a function of the inner
  binaries eccentricity and mass ratio. Now shown as crosses are the
  locations of the inner first stable radii, and still shown as dashed
  lines is the fit of \citet{HW99}. Compared to these as solid lines
  is the 4 parameter fit discussed in the text. Colours indicate mass
  ratio and eccentricity as for Figure~\ref{fig:sp_crit_in}.}
\label{fig:sp_i_fit}
\end{figure}


In this section, the numerical simulations of S(AB)-P type orbits are
described and the results presented. The stability of this region
between the inner and outer binary can be determined using grids of
test particles. This is an efficient method to map out stable radii
for different stellar orbital parameters. Since the stellar orbits and
masses constitute a 7x7x7 dimensional space some simplifying
assumptions are needed to make the investigation feasible. First, all
orbits are taken as coplanar. Second, the initial orientation of the
three stars is fixed, so that they start initially aligned all at
pericentre. The effect of these assumptions will be discussed later.
This leaves the masses, eccentricities and semimajor axes of the stars
as free parameters. As the dynamics will scale with certain
combinations of these, the system should be characterised by ratios of
some of them.

Given the discussion in Section~\ref{sec:stats}, the stellar systems
investigated is as follows. The ratio of the masses of the inner
binary stars $m_A/m_B$ is varied from 1 to 2, and the ratio of the
outer binary $m_C/\mb$ from 0.1 to 2.  Note that these are not the
mass ratios defined in ~\citet{HW99}, which are $\mu_B = m_B/\mb$ and
$\mu_C = m_C/(\mb + m_C)$. The mass of the inner binary is fixed at 2
$M_\odot$ and the eccentricities of both orbits range from 0.0 to
0.6. The semimajor axis of the inner binary is varied between 1 and 5
au and the semimajor axis of the outer binary between 20 and 100
au. The effect of changing these six parameters ($\mu_B$, $\mu_C$,
$e_B$, $e_C$, $a_B$ and $a_C$) is the purpose of this investigation.

For each of these stellar configurations, a grid of test particles is
added. Particles in this grid are spaced evenly in radius and
longitude, and all on initially circular orbits. A semimajor axis is
defined to be stable if all particles starting there remain stable
themselves for the length of the simulation. Individual particles are
considered unstable if their orbits are not bound to the barycentre of
the inner binary or if they pass certain radial limits, as per the
definition of the orbital type in Section~\ref{sec:orbits}. When
either of these conditions are met, the particle is removed from the
simulation.  For S(AB)-P orbits the radial limits are the radius of
the inner and outer binaries. In practice, most particles are lost as
their orbits become unbound.

In their study of planetary stability in binary star systems,
\citet{HW99} find inner and outer critical semimajor axes for test
particles in P and S orbits respectively. These are empirically fitted
to a function of the binary's eccentricity and mass ratio
($\mu=m_2/(m_1+m_2)$). For S(AB)-P orbits in a hierarchical triple system,
it is reasonable to expect there to be an inner and outer critical
semimajor axis, primarily controlled by the inner and outer binaries
respectively. If the stars are separated enough to be well
approximated as two decoupled orbits, then these critical semimajor
axes should be similar in form to those found by \citet{HW99}. There
are two obvious ways that the system differs from this simple picture.
First, as the stellar orbits will vary with time, the initial orbital
configuration may not define the maximum extent of instability regions
controlled by parameters such as eccentricity.  However, for the
systems studied here it will be shown that this is not an important
effect as the orbits do not evolve significantly.  Secondly, the
instability zones in binary systems were shown by \citet{MW06} to be
due to overlap between resonances. It may occur then that there are
additional unstable regions in hierarchical triples where the
resonances due to each sub-binary overlap.

In light of this discussion, simulations can be set up to investigate
the outer and inner stability boundaries separately. First, the outer
boundary can be studied by fixing the inner binary as two 1$M_\odot$
stars in a circular 1 au orbit and varying star C as discussed above.
The test particle grids in these cases are modelled on those chosen by
\citet{HW99}, and extend from the radius of the inner binary to half
that of the outer binary. They are spaced evenly in steps of 0.1 au
and there are eight particles at each semimajor axis.  Simulation
lengths are 1 Myr, which corresponds to at least several thousand
outer binary periods. The timesteps used are of the order of a few
days, giving relative energy conservation at the $10^{-7}$ to
$10^{-8}$ level. For an example of the this, see the right most panels
of Figure~\ref{fig:el}, which shows the energy and angular momentum
variation for a case with star C set to 4$M_\odot$ and with an
eccentricity of 0.6. Also shown in this figure is the variation of the
semimajor axis of the outer binary, to demonstrate that even in the
most extreme case the stellar orbits are not evolving.  This is not
too unexpected as the ratio between the stellar orbits is still fairly
high, even in this case. (For comparison, \citet{EK95} claim that the
stable radius for this hierarchical triple is 16 au, and simulations show
that it disintegrates at an initial separation of around 12 au).

The results of these simulations are given in
Figure~\ref{fig:sp_critout_i}, which shows test particle stability for
each mass ratio. It is easy to define an inner (and outer) first (and
last) stable radius respectively as that being the first (and last)
encountered where all particles are stable. This is plotted for each
eccentricity of star C in the figure. In addition, there are unstable
radii within these bounds, and these are plotted as crosses. For
comparison, the fits of \citet{HW99} are also shown as dashed lines.
Note that these are the optimum values and that there are
uncertainties given for their fit.  It can be seen that the stable
regions here have fairly well defined edges, and in fact match up in
most simulations to those of \citet{HW99}. It is expected around the
inner edge to see some odd unstable radii corresponding to the first
$n:1$ mean motion resonances (MMRs) within the region.

The inner edge of the stable region appears mostly constant as the
orbit of star C is altered, as expected. The only exceptions are the
occasional unstable location near this boundary and a general trend at
small separations of the two stellar binaries for the stability zone
to be less than that expected from the trends of the wider triple
cases. It is possible to demonstrate that, apart from these
exceptions, the outer edge scales the outer binaries semimajor axis by
plotting the data scaled to this quantity, as shown in
Figure~\ref{fig:sp_critout_o} for two of the mass ratio cases.

The only unclear case in these simulations is that of $\mu_C=0.09$ and
$e_C=0.2$.  Here, there is a large unstable region that occurs just
before the last stable radius. This effect is due to a small region of
stability that appears as the last stable radius. Test particles
appear to be stable for just about the length of the simulations.
Interestingly, this island appears to be centred on the 5:1 MMR with
star C. Figure~\ref{fig:tpev01} shows a comparison of the evolution of
test particles in this case and that with the same mass ratio but
$e_C=0.0$. These show as a function of initial semimajor axis average
test particle survival time and fraction of surviving particles. Also
shown are the decay rates in each simulation.  From these, it can be
seen that the dynamics in each $a_C$ case scale with this quantity and
time. It is also clear that in the $a_C=20$ au simulation the region
around the 5:1 MMR is unstable with a lifetime just less than 1Myr,
but that as $a_C$ and the dynamical time increase the region is seen
as stable as the test particles are not lost before the end of the
simulation.  In fact, the test particles within this region all have
very high eccentricities at the end of the simulation as compared to
the rest of the stable zone. In the $e_C=0.0$ graphs it can be seen
that the stable region is very clearly defined, with very few
particles surviving for any length of time outside its boundaries.

Apart from this one feature for the $\mu_C=0.09$ case, the simulation
lengths appear to be sufficiently long to describe stability.
\citet{HW99} use an integration length of $10^4$ binary periods,
comparable to those used here for all but the very low mass ratio and
large $a_C$ cases.  They also mention that the stability boundaries do
not change much after a few hundred binary periods. This would
indicate that even for the long binary period cases, the results are
unaffected by this choice, and this is supported by the scaling seen
with semimajor axis in the location of the outer boundary. As shown in
Figure~\ref{fig:tpev01} and discussed above, the edges to the stable
regions are distinct in most cases, and the test particle decay rates
indicate that a stable situation has been reached long before the end
of the simulations. Further support that the simulations times are
sufficient is given by running the $\mu_C=0.09$ and $e_C=0.6$ cases
for 2 Myr. The results from these are almost identical to the original
1 Myr simulations, with only a few additional unstable points
appearing.

Since the outer boundary is well modelled as a function of mass ratio
and eccentricity and scales with $a_C$ for all but the smallest
separations the $a_C = 100$ au data can be plotted and fitted. This
semimajor axis is chosen as it has the most detailed determination of
the critical radii due to the larger number of test particles used. To
provide a better fit additional simulations were run for six more mass
ratios, and the data to be fitted is shown in the left hand panel of
Figure~\ref{fig:sp_crit_100}.  Instead of using the last stable
radius, the stable locations just within any odd unstable radii around
the boundary are taken instead. Following \citet{HW99}, we fit this
boundary with a function of the form
\begin{equation}
\frac{a_{out}}{a_C} = a_1 + a_2 \mu_C +a_3 e_C
                     + a_4 \mu_C e_C + a_5e_C^2                  
                     + a_6\mu_C e_C^2
\label{eq:hw1}
\end{equation}
Performing this fit gives the parameters as shown in the left hand
column of Table~\ref{tab:hwfit1}. Also given for comparison are the
values suggested by \citet{HW99}. This fit has a $\chi^2$ of about
2600 and is in reasonable agreement with their values. Note however
that here a smaller range of mass ratios has been fitted, as it is not
realistic to increase the mass of star C to the point where $\mu_C$ =
0.9. A four parameter fit to the first four terms in
Equation~\ref{eq:hw1} gives a simpler fit, also shown in the table.
The $\chi^2$ value of this fit is higher at 3700 but still reasonably
models the data, and this fit is shown in the middle and right panels
of Figure~\ref{fig:sp_crit_100}. Note that the $\chi^2$ value is
calculated using the standard formula
\begin{equation}
\chi^2 = \sum \frac{(y_i - y_{obs_i})^2}{\sigma_i^2}
\end{equation}
where $y_i$ is the fitted value of the function, $y_{obs_i}$ the
observed value and $\sigma_i$ the error this value. Since the grid
size here is 0.1 au the stability boundary cannot be located to any
greater accuracy and hence this was assigned as the uncertainty in
each measurement. The large $\chi^2$ values obtained for the fits
reflect the complex nature of a boundary that is not easily fitted
with such a simple function. However, as can be seen from the figures,
it is still a useful approximation for quickly determining the
stability properties of a system.

As mentioned above, the inner stability boundary seems to be
unaffected by changes in the orbit of the outer star except for at
small separations of the two binary orbits. In this case, the stable
region is smaller than expected and there are also far more isolated
unstable radii, especially as the mass of star C increases.
Figure~\ref{fig:sp_crit_detail} shows effect in more detail for the
cases of $\mu_C = 0.09$, 0.60 and 0.67. Here several more semimajor
axes for the outer star have been studied for various eccentricities.
Although there seems to be a linear overlay of the two different
boundaries there does seem to be a slight effect around the point
where they overlap, becoming more pronounced as the mass ratio and
eccentricity increases. There also seems to be a definite unstable
island in the centre of the region, most noticeable in the $e_C = 0.4$
cases. This moves outwards as the position of the outer star is
increased, and must be a resonant feature.  Note that if the mass of
star C is set to zero for the highest mass ratio and eccentricity case
then the stable region has a clearly defined inner radius at 2.3 au
and there are no unstable particles beyond this boundary, indicating
that the structure seen in these graphs is a direct result of the
combined effects of all three stars. Whether this is due to resonance
overlap is unknown.

\begin{table}
\centering
\caption{
  Fitted parameters for Equation~\ref{eq:hw1}, the outer stability edge, 
  compared to those of \citet{HW99}. The first column shows the results 
  of a 6 parameter fit to the data and the middle column the results of 
  a 4 parameter fit.}
\begin{tabular}{lccccccccc}
Parameter& \multicolumn{3}{c}{This work} & \multicolumn{3}{c}{This work} & \multicolumn{3}{c}{\citet{HW99}}\\
\hline
$a_1$ &  0.477 &$\pm$& 0.001  &  0.466 &$\pm$& 0.001 &  0.464 &$\pm$& 0.006 \\
$a_2$ & -0.412 &$\pm$& 0.002  & -0.392 &$\pm$& 0.001 & -0.380 &$\pm$& 0.010 \\
$a_3$ & -0.708 &$\pm$& 0.006  & -0.542 &$\pm$& 0.002 & -0.631 &$\pm$& 0.034 \\
$a_4$ &  0.794 &$\pm$& 0.012  &  0.494 &$\pm$& 0.004 &  0.586 &$\pm$& 0.061 \\
$a_5$ &  0.276 &$\pm$& 0.009  &        & --  &       &  0.150 &$\pm$& 0.041 \\
$a_6$ & -0.500 &$\pm$& 0.020  &        & --  &       & -0.198 &$\pm$& 0.074 \\
\end{tabular}
\label{tab:hwfit1}
\end{table}

\begin{table}
\centering
\caption{
  Fitted parameters for Equation~\ref{eq:hw2}, the inner stability edge, 
  compared to those of \citet{HW99}. The first column shows the results 
  of a 7 parameter fit,and the second column the results of a 4 
  parameter fit.}
\begin{tabular}{lccccccccc}
Parameter& \multicolumn{3}{c}{This work} & \multicolumn{3}{c}{This work} & \multicolumn{3}{c}{\citet{HW99}}\\
\hline
$a_1$ &  3.45 &$\pm$&  1.10  &  2.92  &$\pm$&  0.12  &  1.60  &$\pm$& 0.04  \\
$a_2$ &  9.94 &$\pm$&  1.81  &  4.21  &$\pm$&  0.24  &  5.10  &$\pm$& 0.05  \\
$a_3$ & -6.95 &$\pm$&  1.47  & -2.67  &$\pm$&  0.38  & -2.22  &$\pm$& 0.11  \\
$a_4$ & -5.43 &$\pm$&  5.27  & -1.55  &$\pm$&  0.29  &  4.12  &$\pm$& 0.09  \\
$a_5$ &-14.09 &$\pm$&  4.42  &        & --  &        & -4.27  &$\pm$& 0.17  \\
$a_6$ &  6.22 &$\pm$&  6.26  &        & --  &        & -5.09  &$\pm$& 0.11  \\
$a_7$ & 25.46 &$\pm$&  8.51  &        & --  &        &  4.61  &$\pm$& 0.36  \\
\end{tabular}
\label{tab:hwfit2}
\end{table}

A more detailed investigation of the inner boundary can be carried out
in a similar manner to the outer edge, by fixing star C at 1 $M_\odot$
and in a circular 50 au orbit and varying the inner binaries mass
ratio and eccentricity, as discussed earlier. The radius of the inner
pair was kept at 1 au and their total mass as 2 $M_\odot$. The
eccentricity of this binary was varied from 0.0 to 0.6 in steps of 0.2
again, and the mass ratio $m_A/m_B$ varied from 1.0 to 2.0 in steps of
0.1. Figure \ref{fig:sp_crit_in} shows the results from these
simulations, with the locations of the first and last stable radii
plotted as functions of the inner binaries eccentricity $e_B$ and mass
ratio $\mu_B$. Also plotted again is the fit given by \citet{HW99} to
the critical radius for P type planets in binary systems.  As
expected, the outer stability boundary seems unaffected by changes to
the inner binary pair.

There are some unstable points between the two stability boundaries,
most notably around 3.2 au for the simulations with $e_B=0.0$.
\citet{HW99} find unstable islands appearing at the first $n:1$ beyond
the critical semimajor axis in this configuration. However, the
location of these unstable radii here is well beyond the first $n:1$
MMR in the stable region. In addition, running the $\mu_B=0.34$ and
$e_B=0.0$ case without star C reveals that these locations are now
stable. This would indicate that this is again an effect due to the
combination of all three stars. Because of this the inner edges location
is simply taken as the same as the first stable radius.

The location of this boundary appears to be a very weak function of
mass ratio and an approximately linear function of eccentricity, and
also agrees well with the predictions of \citet{HW99}. For this
boundary, they fit a function of the form
\begin{equation}
\frac{a_{in}}{a_B} = a_1 + a_2 e_B + a_3 e_B^2 + a_4 \mu_B 
                      + a_5 e_B \mu_B + a_6 \mu_B^2 
                      + a_7 e_B^2 \mu_B^2
\label{eq:hw2}
\end{equation}
where the constants are given in the third column of Table
\ref{tab:hwfit2}. The terms up to fourth order are included to fit a
variation in the position of the critical radius at smaller mass
ratios than those considered here. Fitting this function to the
results here does not produce well determined coefficients despite a
low $\chi^2$ value of about 29, as shown in the first column of the
table. In fact, a better solution is obtained by only including the
first four terms, as shown in the middle column of the table. The
$\chi^2$ value for this fit is slightly higher at about 42.  This
second fit is plotted in Figure~\ref{fig:sp_i_fit}, and seems to
describe the data well, although it seems to slightly overestimate the
size of the stable region. Despite the smaller $\chi^2$ values here
the fitted parameters are less well determined than those for the
outer edge. This is a reflection on the relative size of the test
particle grid spacing compared to the size of the inner binaries
orbit. Here the boundary is determined to within a tenth of the
relative separation of the stars, while for the outer edge it was
determined to within a thousandth of the size of the binary's orbit.


\begin{figure}
\centering
\includegraphics[width=\textwidth]{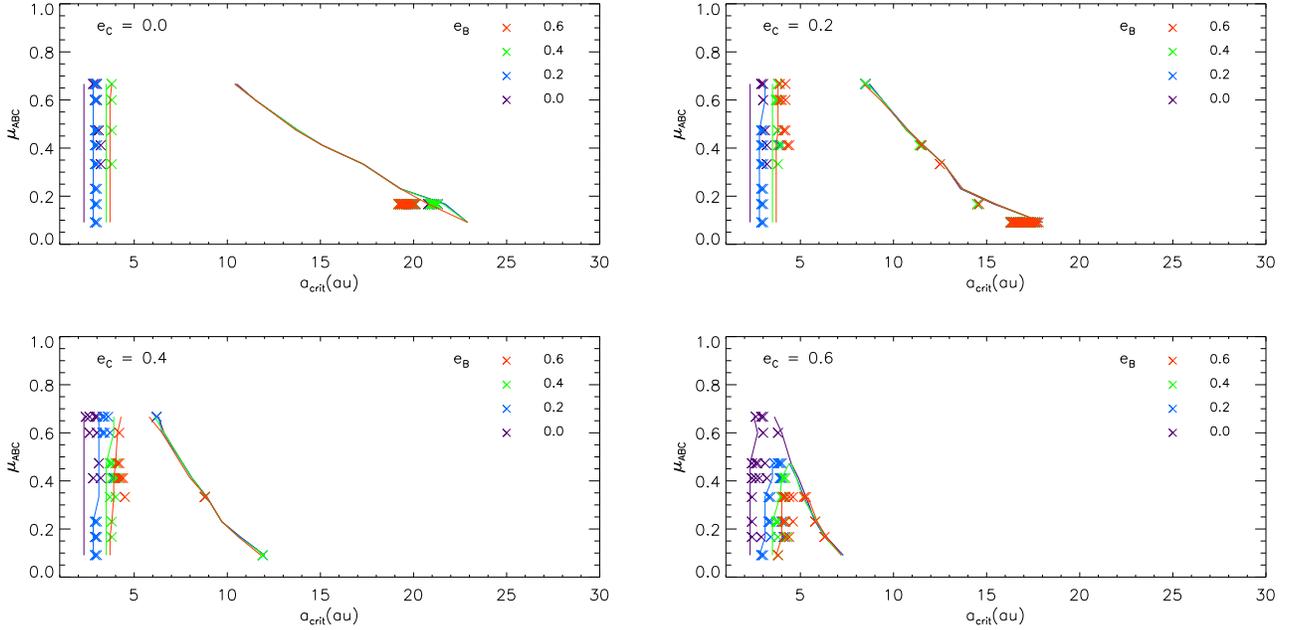}
\caption{
  The stability radii for S(AB)-P type test particles as a function of
  outer binary semimajor axis and mass ratio for different
  eccentricities of both stellar orbits. The colours indicate
  eccentricity of star B and the panels are labelled with the
  eccentricity of star C. Solid lines show the first and last fully
  stable radii, and crosses show any unstable locations within this
  annulus.  }
\label{fig:eccentricity}
\end{figure}



\begin{figure}
\centering
\includegraphics[width=\textwidth]{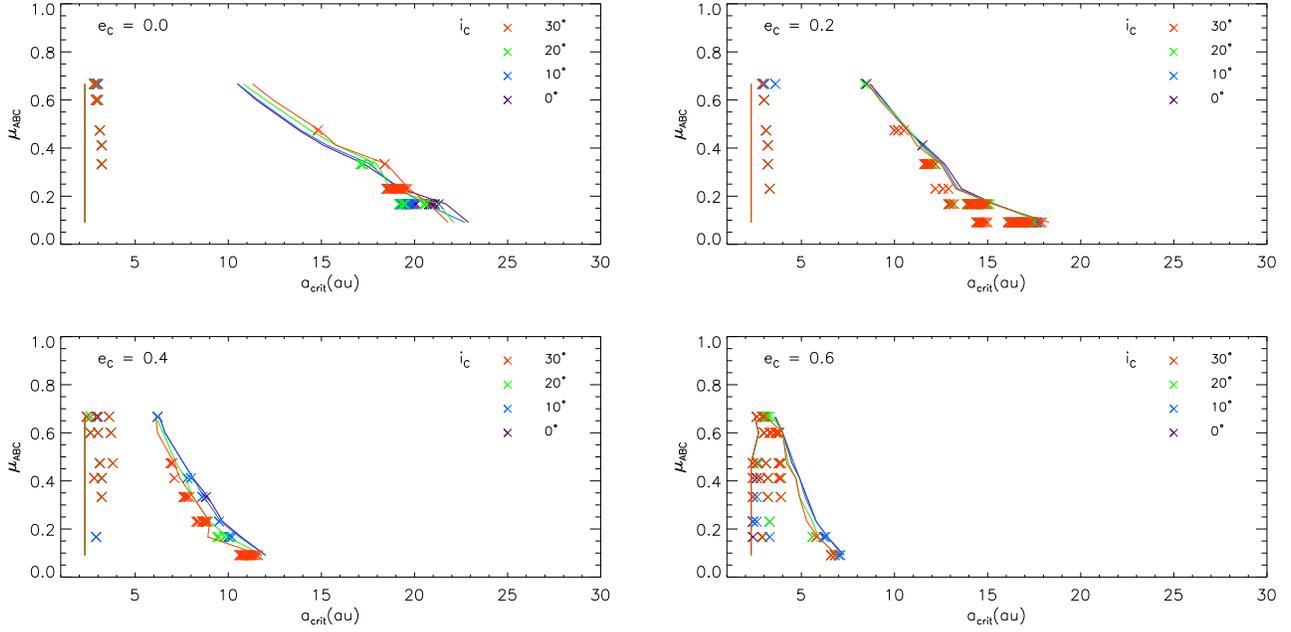}
\caption{
  The stability radii for S(AB)-P type test particles as a function of
  outer binary semimajor axis and mass ratio for different
  eccentricities and inclinations of star C. The colours indicate
  inclination and the panels are labelled with the eccentricity. Solid
  lines show the first and last fully stable radii, and crosses show
  any unstable locations within this annulus.  }
\label{fig:inclination}
\end{figure}


The parameter space investigated so far is somewhat limited. The stars
orbital longitudes have been ignored, assumed to be a minor influence
on stability, the sets of simulations have always kept one star on a
circular orbit, and all objects have been taken as coplanar. Brief
investigations of these three extensions to the parameter space can be
made.

Firstly, the assumption that the initial longitudes of the stellar
orbits has little effect on the stability boundaries was tested by
running the inner edge simulations with mass ratios $\mu_B = $0.33 to
0.40 with a different initial longitude of the inner binary pair. The
results from these simulations were almost identical to those of the
initial set, providing some evidence that this assumption is valid.

Next, the effect of both stars having non-circular orbits was
investigated. The semimajor axes of the inner and outer binary were
kept at 1 and 50 au, and the eccentricities of both varied in steps of
0.2 from 0.0 to 0.6. The mass ratio of the outer binary only was
varied as before. The results of these simulations are shown in
Figure~\ref{fig:eccentricity}. Each panel shows the location of the
stability radii and unstable points as for all values of $e_B$ and one
given value of $e_C$. There is almost no change in the position of the
outer stability boundary, but as both stellar eccentricities and the
mass ratio increase, the inner boundary starts to move outwards. The
stellar eccentricities are still not varying to any significant extent
and the additional instability must be due to the combined
perturbations of all three stars.

Lastly, the effect of inclination was studied. The semimajor axes were
kept as 1 and 50 au, $e_B$ was set to 0, and the outer mass ratio and
eccentricity varied as before. Sets of simulations where then run for
inclinations of the outer star of $10^\circ$, $20^\circ$ and
$30^\circ$. The test particles were kept coplanar with the inner
binary.  This is a rather limited investigation, as any dependence on
the longitude of ascending node has been ignored, and only a small
range of inclinations included. However, higher inclinations will be
subject to the Kozai instability, causing large variations in the
stellar orbits, which is expected to rapidly destabilise test
particles. In these simulations the stellar mutual inclination remains
fairly constant and their orbits are stable.
Figure~\ref{fig:inclination} shows the stability boundaries for these
simulations, each panel comparing the different inclination results
for a different value of $e_C$. There is little change in the inner
boundary but the outer edge moves somewhat. Interestingly, for the
$e_C=0$ case higher inclinations are more stable.  If the test
particles are started instead coplanar with the outer binary the
stability is similar, although not identical. These results are
consistent with the conclusions of \citet{PFD03}, who show that
inclination is not a significant effect on the stability of P type
planets in binary system.


\section{Conclusion}
\label{sec:conclusion}

The main achievement of this paper is the formulation of a symplectic
integrator algorithm suitable for hierarchical triple systems. This
extends the algorithm for binary systems presented by \citet{CQDL02}.
The positions of the stars are followed in hierarchical Jacobi
coordinates, whilst the planets are referenced purely to their
primary. Each of the five distinct cases, namely circumtriple orbits,
circumbinary orbits and circumstellar orbits around each of the stars
in the hierarchical triple, requires a different splitting of the
Hamiltonian and hence a different formulation of the symplectic
integration algorithm. Here, we have given the mathematical details
for each of the five cases, and presented a working code that
implements the algorithm.

As an application, a survey of the stability zones for circumbinary
planets in hierarchical triples is presented. Here, the planet orbits
an inner binary, with a more distant companion star completing the
stellar triple. Using a set of numerical simulations, we found the
extent of the stable zone which can support long-lived planetary
orbits and provided fits to the inner and outer edges. The effect of
low inclination on this boundary is minimal. A reasonable first
approximation to a behaviour of a hierarchical triple is to regard it
as a superposition of the dynamics of the inner binary and a
pseudo-binary consisting of the outer star and a point mass
approximation to the inner binary. If it is considered as two
decoupled binary systems, then the earlier work of Holman \& Wiegert
(1999) on binaries is applicable to triples, except in the cases of
high eccentricities and close or massive stars.

The implication here is that the addition of a stable third star does
not distort the original binary stability boundaries. As mentioned,
\citet{MW06} have shown that overlapping sub-resonances are the cause
of the boundary in the binary case. It is reasonable to expect that in
triples the same process is responsible, and the similarities between
the binary and triple results support this theory. It is also expected
that there is a regime in which the resonances from each sub-binary
start to overlap as well, further destabilising the test particles.
Evidence of this is the deviation from the binary results when the
stars are close, massive and very eccentric, when resonances would be
both stronger and wider. Since the parameter space investigated was
chosen to reflect the observed systems it would seem to be a
reflection on the characteristics of known triple stars that most lie
in the decoupled regime. The relatively constant nature of the stellar
orbits in the simulations is however a consequence of the test
particle orbits being destabilised long before the stars are close
enough to interact. By extension of all these arguments, it is
expected that the binary criteria can be used to fairly accurately
predict the stability zones in any hierarchical stellar system, no
matter the number of stars.

\begin{figure}
\centering
\includegraphics[width=\textwidth]{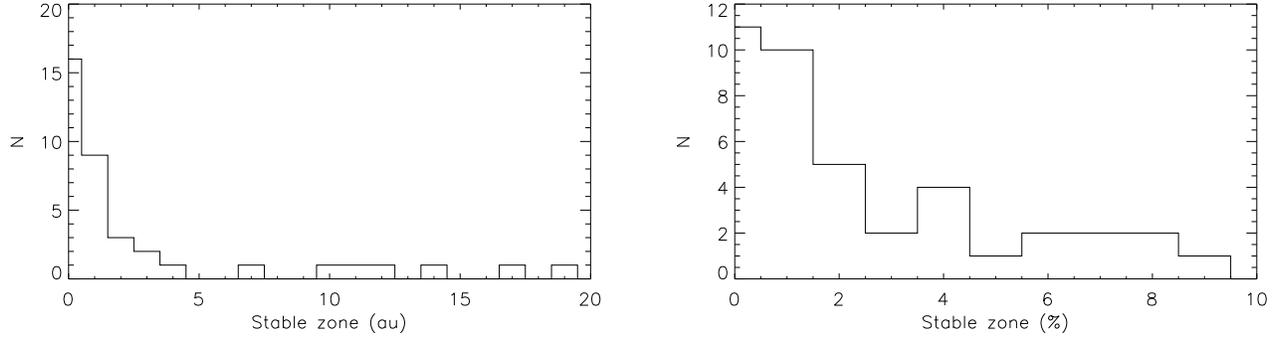}
\caption{
  Widths of the circumbinary stability zones for known triple systems
  in au and as a percentage of the area between the two sub-binaries,
  calculated using the four parameter fits derived in
  Section~\ref{sec:sp}. Note that using the criteria of \citet{HW99}
  gives almost identical results.}
\label{fig:zone}
\end{figure}

The results presented here can be used to estimate the number of known
hierarchical triple systems that could harbour S(AB)-P planets.
\citet{To97} lists 54 systems with semimajor axis, eccentricity and
masses for both the inner and outer components. The mutual
inclinations of most are not well known, but there are nine systems
listed for which this angle can be determined.  For five of these it
is less than $15^\circ$, two are around $40^\circ$ and two are
retrograde. Although a small sample it suggests that there are systems
that fall within the low inclination regime investigated here.  Using
the criteria of \cite{HW99} and those found here for the position of
the inner and outer critical semimajor axis the size of the coplanar
stable region for each of these triples can be calculated.  This can
be considered an upper limit, since it is likely that very
non-coplanar systems and those with significant eccentricities for
both binary components will further destabilise planets.  Of the 54
systems 13 are completely unstable to circumbinary planets according
to the four parameter fits (compared to 11 using \citet{HW99}'s
criteria). Figure \ref{fig:zone} shows a plot of the width of the
stable region for the remaining systems.  Interestingly, the majority
seem to have very small stable zones, with 16 smaller than an au.
Whether this is a feature of triple systems, or an observation bias is
not apparent.  It does indicate though that circumbinary planets are
unlikely to exist in at least 50 \% of observable systems.


\appendix

\section{Appendix}
\label{sec:appendix}

In this appendix are details of the symplectic integration method
derived in Section~\ref{sec:maths} for the orbital types P, S(A), S(B)
and S(C). The split Hamiltonian in each case is presented, and the
evolution under these is readily obtainable using the method given in
Section~\ref{sec:maths}. Also presented are details of the testing of
the implementation of this symplectic integrator. Note that throughout
this appendix $\mt$ is the total system mass and $\mb$ the total inner
binary mass, including planets where relevant.

For S(A) orbits, the Hamiltonian is
\begin{eqnarray}
H &=& \Hk + \Hi + \Hj \lf
\Hk &=& \frac{\bP_{B}^{2}}{2\mu_{bin}} + \frac{\bP_{C}^{2}}{2\mu_{tot}}
               - \frac{G\mu_{bin}\mb}{R_B} - \frac{G\mu_{tot}m_{tot}}{R_C}    
               + \Z \left( \frac{\bP_{i}^2}{2m_i}-\frac{Gm_{A}m_i}{R_i}  \right)\lf
\Hi &=& -\Z \sum_{j>i} \frac{Gm_im_j}{R_{ij}}+Gm_Am_B\Bigg(\frac{1}{R_B}-\frac{1}{|\bX_B+\bs|}\Bigg)
\lf
& &{} +G \Z m_i\Bigg(\frac{m_B}{R_B}-\frac{m_B}{|\bX_B-\bX_i+\bs|}   
                   -\frac{m_C\mb}{|\mb\bX_i-m_B\bX_B-\mb\bX_C-\mb\bs|}\Bigg)\lf
& &{} +Gm_C\mb\Bigg(\frac{1}{R_C}-\frac{m_A}{|m_B\bX_B+\mb\bX_C+\mb\bs|}   
                   -\frac{m_B}{|\mb\bX_C-\map\bX_B|}\Bigg)\lf
\Hj &=& \frac{|\ZP|^2}{2m_A}
\end{eqnarray}
where $\map$ is the mass of star A and its planets, $\mu_{bin}$ is the
reduced mass of the inner binary, including the planets, $\mu_{tot}$
is the reduced mass of the outer binary and $\bs = \ZmX/\map$.

For S(B) type planets only, the equations are
\begin{eqnarray}
H &=& \Hk + \Hi + \Hj \lf
\Hk &=& \frac{\bP_{B}^{2}}{2\mu_{bin}} + \frac{\bP_{C}^{2}}{2\mu_{tot}}
               - \frac{G\mu_{bin} \mtc }{R_B} - \frac{G\mu_{tot}\mt}{R_C}    
               + \Z \left( \frac{\bP_{i}^2}{2m_i}-\frac{Gm_{B}m_i}{R_i} \right)\lf
\Hi &=& -\Z \sum_{j>i} \frac{Gm_im_j}{R_{ij}}    
            + Gm_A\Bigg( \frac{\mbp}{R_B} - \frac{m_B}{|\bX_B - \bs|} - \frac{m_C}{|r\bX_B + \bX_C|} \Bigg)\lf
& &{}       + Gm_C\Bigg(\frac{ \mtc }{R_C} - \frac{m_B}{|q\bX_B - \bX_C -\bs|}   \Bigg)   
            - G\Z m_i \Bigg(\frac{m_A}{|\bX_i+\bX_B-\bs|} + \frac{m_C}{|\bX_i+q\bX_B-\bX_C-\bs|} \Bigg)\lf
\Hj &=& \frac{|\ZP|^2}{2 m_B}
\end{eqnarray}
where $\mbp$ is the mass of star B and its planets, $\mu_{bin}$ is the
reduced mass of the inner binary, including the planets and $\mu_{tot}$
is the reduced mass of the outer binary. Here $\bs = \ZmX/\mbp$ now,
and $q = m_A/\mtc$ and $r= \mbp/\mtc$.

The corresponding Hamiltonian for S(C) type planets is
\begin{eqnarray}
H &=& \Hk + \Hi + \Hj \lf
\Hk &=& \frac{\bP_{B}^{2}}{2\mu_{bin}} + \frac{\bP_{C}^{2}}{2\mu_{tri}}
               - \frac{Gm_Am_B}{R_B} - \frac{G\mcp\mb}{R_C}    
               + \Z \left( \frac{\bP_{i}^2}{2m_i}-\frac{Gm_{C}m_i}{R_i} \right) \lf 
\Hi &=& -\Z \sum_{j>i} \frac{Gm_im_j}{R_{ij}} +\frac{G\mcp\mb}{R_C}      
            -Gm_C\Bigg(\frac{m_A}{|\bX_C+r\bX_B-\bs|} + \frac{m_B}{|\bX_C-q\bX_B-\bs|}  \Bigg)\lf     
& &{}       -G\Z m_i\Bigg( \frac{m_A}{|\bX_i + \bX_C+r\bX_B-\bs|}    
                     + \frac{m_B}{|\bX_i+\bX_C-q\bX_B-\bs|}\Bigg)\lf
\Hj &=& \frac{|\ZP|^2}{2m_C}
\end{eqnarray}
where $\mu_{bin}$ is the reduced mass of the inner binary stars and
$\mu_{tri}$ is the reduced mass of the outer binary, including the
planets. Here now $\bs = \ZmX/\mcp $, $q = m_A/\mb $, $ r = m_B/\mb$
and $\mcp$ is the mass of star C and its planets.

Finally, for P orbits, the equations are
\begin{eqnarray}
H \quad \,\,&=& \Hk + \Hi + \Hj \lf
\Hk &=& \frac{\bP_{B}^{2}}{2\mu_{bin}} + \frac{\bP_{C}^{2}}{2\mu_{tri}}
               - \frac{G\mu_{bin}\mb}{R_B} - \frac{G\mu_{tri}m_{tri}}{R_C}    
               + \Z \left( \frac{\bP_{i}^2}{2m_i}-\frac{Gm_{tri}m_i}{R_i} \right)\lf
\Hi &=& -\Z \sum_{j>i} \frac{Gm_im_j}{R_{ij}}    
      +Gm_C\mb\Bigg(\frac{1}{R_C} - \frac{m_A}{|m_B\bX_B+\mb\bX_C|}   
                          -\frac{m_B}{|m_A\bX_B-\mb\bX_C|}\Bigg)\lf
& &{} -G\mb\Z m_i\Bigg(\frac{m_A}{|\mb\bX_i+m_B\bX_B+\mu_{tri}\bX_C|}   
                                       +\frac{m_B}{|\mb\bX_i-m_A\bX_B+\mu_{tri}\bX_C|}\Bigg)\lf
& &{} +Gm_{tri}\Z m_i\Bigg(\frac{1}{R_i}-\frac{m_C}{|m_{tri}\bX_i-\mb\bX_C|}\Bigg)\lf
\Hj &=& \frac{|\ZP|^2}{2m_{tri}}
\end{eqnarray}
where $\mu_{bin}$ is as before, $\mu_{tri}$ is the reduced mass of the
outer binary's orbit and $m_{tri}$ is the mass of all three stars.

These differing orbital cases are all dealt with separately at present.
To test the method and its implementation comparisons of the
simulations of systems of various levels of complexity were made with
the literature. As the equations given above allow $m_C$ to be set to
zero, the conservation of the Jacobi constant of test particles in the
Circular Restricted Three Body Problem was calculated and compared to
that given by \citet{CQDL02} for their binary systems.  The evolution
of various triple star system was compared to studies of hierarchical
triples in the literature, for example those given in \citet{Be03}.
Simulations of discs of test particles were also compared to those
given by \citet{Be03} for circumbinary discs. More complex systems of
planets and test particles were also compared to the results of a
standard Bulirsch-Stoer integrator \citep{NR}. In all cases excellent
agreement was found.


\label{lastpage}


\begin{thebibliography}{}

\bibitem[Beust(2003)]{Be03} 
Beust, H.\ 2003, A\&A, 400, 1129

\bibitem[Broucke(2004)]{Br04} 
Broucke, R.~A.\ 2004, Ann. New York Acad. Sciences, 1019, 408 

\bibitem[Chambers et al.(2002)]{CQDL02} 
Chambers, J.~E., Quintana, E.~V., Duncan, M.~J., \& Lissauer, J.~J.\ 2002, AJ, 123, 2884 

\bibitem[Danby(1988)]{Da88} 
Danby, J.~M.~A.\ 1988, Richmond, Va., U.S.A.~: Willmann-Bell, 1988.~2nd ed.

\bibitem[Desidera \& Barbieri(2007)]{DB07} 
Desidera, S., \& Barbieri, M.\ 2007, A\&A, 462, 345 

\bibitem[Dvorak et al.(2003)]{Dv03} 
Dvorak, R., Pilat-Lohinger, E., Funk, B., \& Freistetter, F.\ 2003, 
A\&A, 398, L1

\bibitem[Dvorak (1986)]{Dv86}
Dvorak, R.\ 1986, A\&A, 167, 379 

\bibitem[Dvorak(1984)]{Dv84}
Dvorak, R.\ 1984, Celestial Mechanics, 34, 369 

\bibitem[Eggenberger et al.(2007)]{eggenb}
Eggenberger A., Udry S., Mazeh T., Segal Y., Mayor M., 2007, A\&A, in press

\bibitem[Eggleton \& Kiseleva(1995)]{EK95} 
Eggleton, P., \& Kiseleva, L.\ 1995, ApJ, 455, 640 

\bibitem[Haghighipour(2006)]{Ha06} 
Haghighipour, N.\ 2006, ApJ, 644, 543 

\bibitem[Holman \& Wiegert(1999)]{HW99}
Holman, M.~J., \& Wiegert, P.~A.\ 1999, AJ, 117, 621 

\bibitem[Konacki(2005)]{Ko05} 
Konacki, M.\ 2005, Nature, 436, 230 

\bibitem[Mudryk \& Wu(2006)]{MW06} 
Mudryk, L.~R., \& Wu, Y.\ 2006, ApJ, 639, 423 

\bibitem[Murray \& Dermott(2000)]{MD00} Murray C.~D., 
Dermott S.~F. 2000, Solar System Dynamics, Cambridge University Press

\bibitem[Pilat-Lohinger et al.(2003)]{PFD03} 
Pilat-Lohinger, E., Funk, B., \& Dvorak, R.\ 2003, A\&A, 400, 1085 

\bibitem[Press et al.(2002)]{NR}
Press, W.~H., Teukolsky, S.~A., Vetterling, W.~T., \& Flannery, B.~P.\ 2002,
Numerical recipes in C++ : the art of scientific computing, Cambridge University Press

\bibitem[Szebehely(1977)]{Sz77} 
Szebehely, V.\ 1977, Revista Mexicana de Astronomia y Astrofisica, 
vol.~ 3, 3, 145 

\bibitem[Szebehely(1967)]{Sz67} 
Szebehely, V.\ 1967, Theory of Orbits, Chapter 9, New York: Academic Press

\bibitem[Tokovinin(1997)]{To97} 
Tokovinin, A.~A.\ 1997, A\&AS, 124, 75

\bibitem[Verrier(in prep.)]{Ve08}
Verrier, P.~E. \ in prep., PhD Thesis

\bibitem[Verrier \& Evans(2006)]{VE06} 
Verrier, P.~E., \& Evans, N.~W.\ 2006, MNRAS, 368, 1599 

\bibitem[Wisdom \& Holman(1991)]{WH91} 
Wisdom, J., \& Holman, M.\ 1991, AJ, 102, 1528 


\end{thebibliography}
\end{document}